\documentclass[12pt]{article}
\usepackage{epsfig,graphicx}
\usepackage{ch2005}

\def\beq{\begin{equation}}
\def\eeq#1{\label{#1}\end{equation}}
\def\eeqn{\end{equation}}

\def\beqa{\begin{eqnarray}}
\def\eeqa#1{\label{#1}\end{eqnarray}}
\def\eeqan{\end{eqnarray}}

\def\Dslash{\not{\hbox{\kern-4pt $D$}}}
\def\dslash{\not{\hbox{\kern-2pt $\del$}}}

\newcommand{\tev}{\ensuremath{\mathrm{\,Te\kern -0.1em V}}\xspace}
\newcommand{\gev}{\ensuremath{\mathrm{\,Ge\kern -0.1em V}}\xspace}
\newcommand{\mev}{\ensuremath{\mathrm{\,Me\kern -0.1em V}}\xspace}
\newcommand{\kev}{\ensuremath{\mathrm{\,ke\kern -0.1em V}}\xspace}
\newcommand{\ev}{\ensuremath{\mathrm{\,e\kern -0.1em V}}\xspace}
\newcommand{\gevc}{\ensuremath{{\mathrm{\,Ge\kern -0.1em V\!/}c}}\xspace}
\newcommand{\mevc}{\ensuremath{{\mathrm{\,Me\kern -0.1em V\!/}c}}\xspace}
\newcommand{\gevcc}{\ensuremath{{\mathrm{\,Ge\kern -0.1em V\!/}c^2}}\xspace}
\newcommand{\mevcc}{\ensuremath{{\mathrm{\,Me\kern -0.1em V\!/}c^2}}\xspace}

\def\mus  {\ensuremath{\rm \,\mus}\xspace}

\def\mus        {\ensuremath{\,\mu{\rm s}}\xspace}    

\begin{document}


\Title{The Galactic Center at High Energies}
\bigskip


%
\label{GoldwurmStart}

%
\author{ Andrea Goldwurm\index{Goldwurm, A.} }

%
\address{Service d'Astrophysique - CEA Saclay \\
F-91191 Gif sur Yvette Cedex, 
France \\
}

\makeauthor\abstracts{
I present here a review on the high energy phenomena occurring 
in the Galactic Center region, and report in particular on the results 
obtained from recent X-ray and gamma-ray observations.
}

\section{The center of the Galaxy}
The galactic center (GC) is a very dense and complex sky region of 
approximately 600 pc size ($\sim$ 4$^\circ$ in projection) 
where a number of interesting high-energy phenomena take place.
Located at about 8 kpc distance it hosts the nearest 
super massive black hole (SMBH)
surrounded by a variety of objects which interact with each other.
The high energy processes generated in this extreme environment 
are possibly common to other galactic nuclei, 
and, given its proximity, the GC rapresents a unique 
laboratory for modern astronomy.
Totally obscured in the optical wavelengths by the
galactic plane, the GC is mainly observed from radio to infrared (IR) and 
again at high energies.
Some important results have been  
recently obtained with the new generation 
of X-ray and gamma-ray observatories: 
Chandra, XMM-Newton, INTEGRAL and HESS. 
After a short introduction on the GC region 
(see also the reviews \cite{mez96,mor96,gol01,melfal01}). 
I will summarize and discuss some of these results.

\begin{figure}[htb]
\begin{center}
\epsfig{file=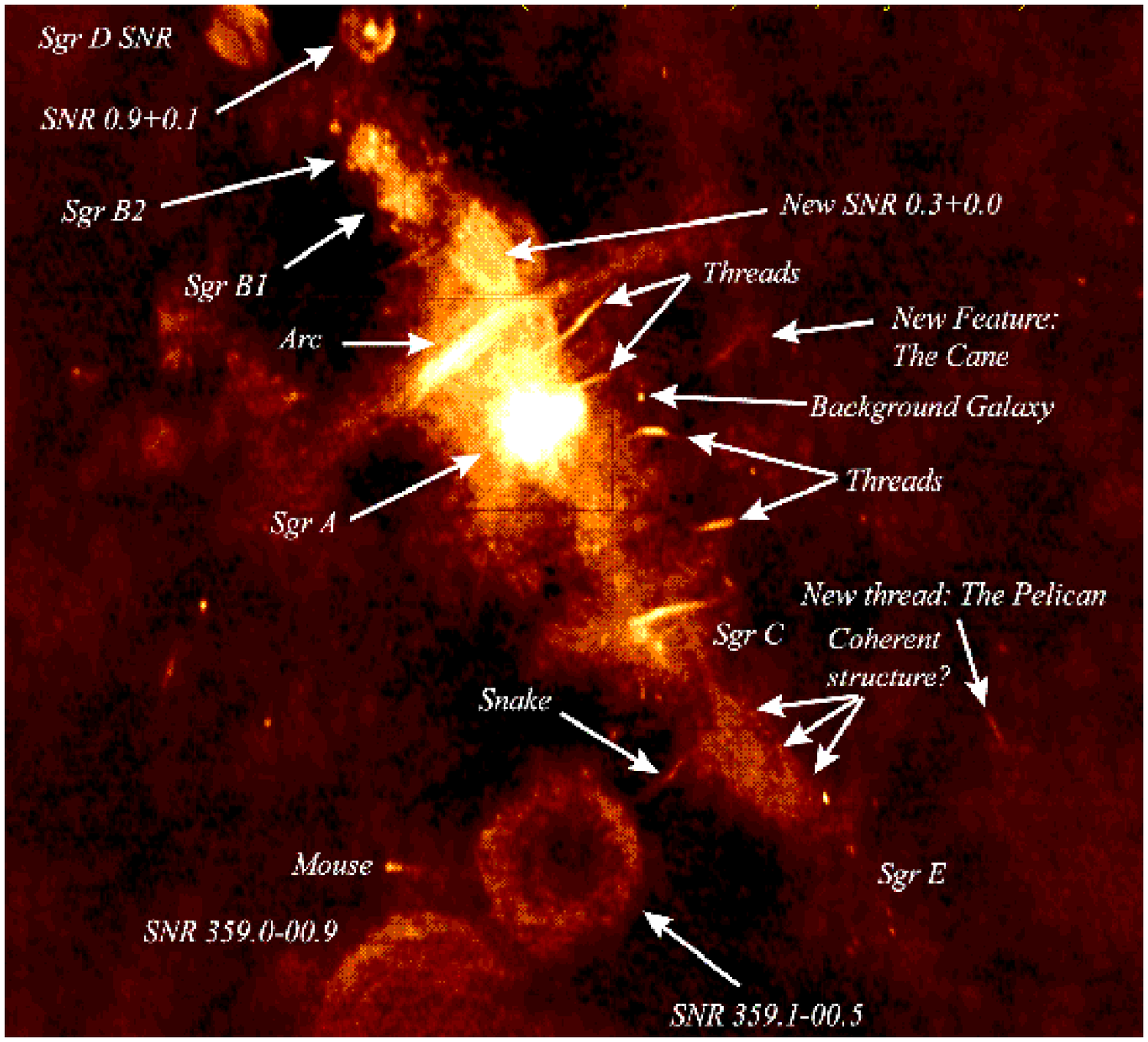, width=5.5cm} ~~~~
\epsfig{file=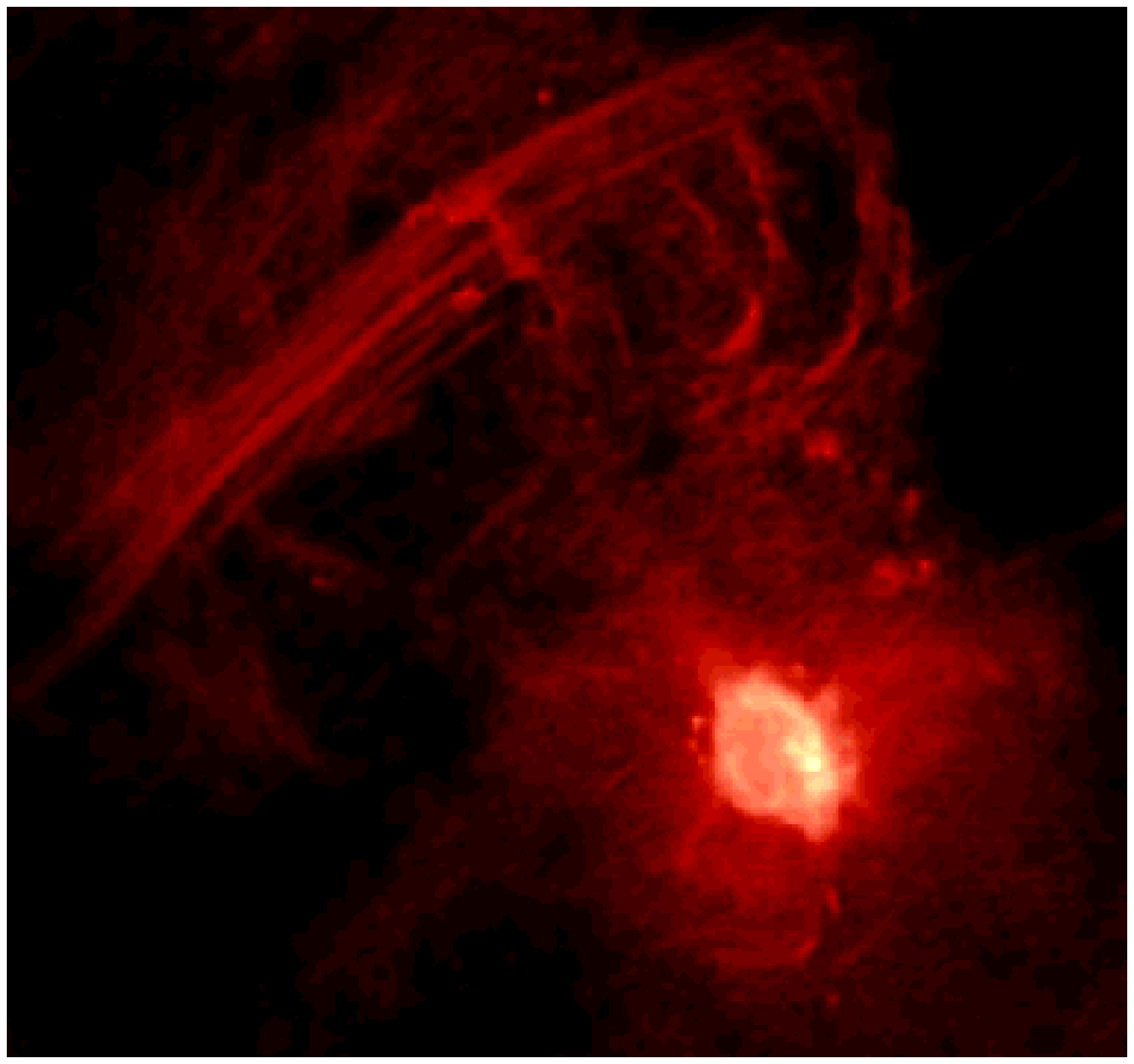,width=5.3cm}
\caption{The Galactic center region seen with the VLA at 90~cm (La Rosa et al. 2000) 
(left) and the Sgr A complex with the radio arc seen with the VLA at 20~cm 
(Yusef-Zadeh et al. 2002, ApJ,570,665) (right).
}
\label{fig:Goldwurm-fig1}
\end{center}
\end{figure}

The beautiful radio image of the GC obtained with the VLA at 90~cm 
(Fig.~\ref{fig:Goldwurm-fig1} left) 
\cite{lar00} shows all the complexity of this region. 
The GC contains $\sim$ 10$\%$ of the galactic interstellar medium (ISM), concentrated
in the dense giant molecular clouds (MC) like Sgr B, Sgr C and those
of the Sgr A radio and molecular complex. 
Shell-like supernova remnants (SNR) heat the ISM with their expanding shells
(e.g. G~359.1-005) which appear sometimes interacting with the MCs.
While several other non-thermal filaments demonstrate the presence of accelareted 
particles spiralling the strong magnetic fields ($\sim$~1~mG) of the region 
(like the large structure known as the {\it radio arc}, Fig.~\ref{fig:Goldwurm-fig1} right),
other structures have thermal radio spectra and are in fact
HII regions ionized by closeby hot and young star clusters.
The central 30 pc are dominated by the Sgr A complex
(Fig.~\ref{fig:Goldwurm-fig1} right), formed by few MCs
(M-0.02-0.07, M-0.13-0.06) an expanding SNR, Sgr~A~East,
and a central HII region surrounding the bright compact radio source Sgr~A$^*$, 
the radio manifestation of the central SMBH.
Sgr~A~East is a non-thermal radio source composed by a diffuse emission
of triangular shape and an inner oval shell (7 pc $\times$ 9 pc i.e.
3$'~\times$~4$'$) centered about 50$''$ ($\approx$ 2~pc) west of Sgr~A$^*$.
The shell appears in expansion, compressing the molecular cloud M-0.02-0.07
and probably creating the string of 4 HII regions and the OH masers 
observed around the shell.
The first estimates of shell energy
were well above the typical release of a SN, and it was proposed
that Sgr A East was the result of 40 SN or of the esplosive 
tidal disruption of a star by the SMBH.  
In the inner regions a rotating molecular ring surrounds Sgr~A~West, 
a thermal diffuse nebula with the characteristic shape of a minispiral,
also rapidly rotating around the compact source Sgr~A$^*$. 
Sgr~A~West is ionized by a cluster of hot young and massive stars, 
centered at about 2$''$ from Sgr~A$^*$ and known as IRS~16 (IRAS source).
Some of these stars emit powerful stellar winds which interact
with the surrounding medium and probably feed the SMBH. 

While the matter dynamics at radial distances $>$ 2 pc
is dominated by the central core of the galactic bulge star cluster, 
the large velocities of gas and stars
observed in the innermost regions must imply the presence
of a massive black hole.
The adaptive optics NIR measures, made with the NTT, the VLT and the Keck
over the last 10-15 yr, 
of velocities and proper motions of the brightest and closest stars 
to Sgr~A$^*$ (the central star cluster) have by now
provided precise orbital parameters for several of them \cite{sch03,eis05,ghe05}).
The derived parameters imply the presence of a dark mass 
of 3-4 10$^6$~M$_{\odot}$ enclosed within a radius $<$ 100~AU. 	
Only a SMBH can explain such densities.
The dynamical center of the central star cluster is coincident
(within 10 mas) with the bright ($\approx$ 1 Jy), compact, 
variable, synchrotron (flat power law spectrum) radio source Sgr~A$^*$.
Since its discovery, 30 years ago, it has been considered 
the counterpart of the massive black hole of the Galaxy. 
The source is linearly polarized at sub-mm frequencies
where the spectrum also present a bump indicating that
the emission becomes optically thin. 
Sgr~A$^*$ proper motion is $<$ 20 km/s and its size, measured
at frequencies of 3~mm where the interstellar scattering is small, 
is of the order of 0.1-0.3~mas, about 15-20~R$_S$, 
where R$_S$ = ${2 G M \over c^2}$ = 10$^{12}$~cm = 0.06~AU
is the Schwarzschild radius for a 3.5~10$^6$~M$_{\odot}$~BH.
%
%
\section{The Galactic Center in the X-ray band}
In the 2-10 keV X-ray band the GC has been deeply monitored by Chandra 
and XMM-Newton. It shows the following components:
few bright sometimes transient X-ray binaries 
probably not associated to the GC (e.g. 1E1740.7-2942, 1E1743.1-2843); 
a large population of weak point-like persistent and transient sources;
a diffuse emission with 3 distinct components, a soft
thermal one (kT $\sim$1 keV, probably SNR heated gas), 
a hot component (kT $\sim$8 keV)
and a nonthermal one characterized by a strong 6.4~keV line.
Several SNR, non-thermal filaments and star clusters are also detected. 
The central 20~pc emission is dominated by the thermal 
emission from Sgr A East, while Sgr A* itself appears very weak.
\subsection{Quiescence and flaring activity of Sgr A*}
A SMBH surrounded by dense environment is an ideal system
to generate accretion powered X-ray emission.
And indeed the first reports of high energy emission from the GC
direction were attributed to the SMBH. However as the
resolution and sensitivity of the high energy telescopes increased 
it was realized that the galactic SMBH is a very weak,
even in the hard X-ray domain (where BH binaries in hard state
emit the bulk of their accretion luminosity) \cite{gol94, gol01}. 
The total (from radio to X-rays) Sgr~A$^*$ luminosity amounts 
to less than 5~10$^{36}$~ergs~s$^{-1}$, i.e. some 10$^{-8}$ 
times the Eddington luminosity of a 3.5~10$^6$~M$_{\odot}$ BH.
Since the IRS16 stellar winds are supposed to feed the BH
at a rate of few 10$^{-4}$~M$_{\odot}$~yr$^{-1}$
which implies accretion luminosities of 0.02~L$_E$,
this led to the development of theories of very inefficient 
accretion flows (e.g. the so called ADAF models) \cite{melfal01}.  
The Chandra observatory in 1999, with its unprecedent 
angular resolution of 0.5$''$
confirmed the very low X-ray luminosity of Sgr~A$^*$ 
(2~10$^{33}$~ergs~s$^{-1}$ in the 2-10~keV band)
but measured a steep spectrum ($\alpha \sim$ 2.5), not
compatible with the ADAF thermal bremsstrahlung models (\cite{bag03}).
One year later Chandra made the dramatic discovery of 
a powerful X-ray flare from Sgr~A$^*$. 
During this event, of a total duration of 3~hr, 
the flux increased by factor 50 to 
reach luminosities of 10$^{35}$~ergs~s$^{-1}$ displaying 
a hard spectral slope ($\alpha \sim$ 1.3) \cite{bag01}.
XMM-Newton confirmed the presence of such bright hard flares
from Sgr~A$^*$ \cite{gol03} and discovered the most powerful
one with an increase factor of 200 and, this time, a significantly 
steeper spectrum ($\alpha \sim$2.5) \cite{por03}.
The flare duration (few hours) and the observed short
time scales variations (200~s) indicate
that the X-ray emission is produced within 20~R$_S$.
This cannot be accounted for by the standard ADAF model 
(for which the bulk of the X-ray emission is produced from the
whole accretion flow starting at the accretion radius) and 
several other models are now considered where non-thermal
emission plays a major role. 
The Liu and Melia model \cite{liumel02,liu04} assumes that
accreting matter circularizes in a small, very hot, magnetized
keplerian disk where quasi relativistic electrons 
produce synchrotron radiation in the sub-mm band and, 
by inverse compton, the steep X-ray spectrum. Flares can
be produced either by sudden increase in accretion rate 
or release of magnetic energy and the 2 different spectral slopes 
can be explained.
Markoff et al. \cite{mar01} locate the main energy release at the base
of a relativistic jet rather than in the accretion disk.
Substantial modification of ADAF models (inclusion of
outflows, convection and non-thermal component) 
were also considered \cite{yua03}. 
The different models can account for the observed spectral 
shapes but they predict different correlations between sub-mm, NIR 
and X-ray fluxes. The multiwavelength observation of Sgr~A$^*$ flares 
could allow to identify the correct model.
\subsection{The diffuse X-ray emission}
The Chandra and XMM-Newton surveys of the GC have 
also provided several new results on the diffuse emission.
The first one is the confirmation that the central few hundred parsecs
are permeated by a 
hard diffuse emission peaked towards the center and extendig
along the plane, as observed by previous instruments 
and in particular by ASCA \cite{koy96}. 
Continuum and line spectra of this emission \cite{par04,mun04} 
(and in particular the strong 6.7 keV of ionized iron)
seem to indicate that it is thermal with temperature of 8~keV. 
Such a hot plasma cannot be confined in the region by the 
gravitational potential, it would escape in $<$ 4~10$^4$~yr 
and its origin is therefore unexplained (but see \cite{belm05}).
Chandra detected 2000 point-like weak sources in the central 17$'$~$\times$~17$'$ 
but this population cannot explain more than 10$\%$ of the diffuse 
emission \cite{mun04}.
Some features of this component (the continuum
is sometimes associated to the 6.4 keV line rather than the 6.7 keV one) 
are difficult to reconcile with a thermal nature and few authors 
have proposed non-thermal origin, i.e. cosmic ray interaction with ISM \cite{val00,yus02},
or effect of SN ejecta in dense regions \cite{byk03}.
Indeed the other distinct component of the GC diffuse emission 
is the 6.4 keV line of neutral or weakly ionized iron, which 
has a different morphology than the 6.7~keV line and is certainly 
due to non-thermal processes, involving reprocessing of external high energy radiation 
or cosmic ray interactions with dense MC.  
\subsection{Sgr B2 and Sgr A East}
The 6.4~keV image of the region shows a very strong peak at the position 
of the Sgr~B2 GMC. 
This was interpreted as fluorescent line due to scattering of 
hard X-ray emission coming from an external source, possibly Sgr~A$^*$
itself \cite{koy96,mur00}. A strong transient outburst of hard X-rays from the SMBH 
occured some 300~yr back would have travelled the distance 
to Sgr B2 illuminating the dense cloud and generating the Fe line
along with hard X-ray emission. 
Such hard ($>$ 10 keV) scattered emission was initially detected 
with GRANAT/ART-P and now with INTEGRAL \cite{rev04}.
The X-ray observations of the Sgr A complex have also demonstrated 
that the bright X-ray source Sgr~A~East is a mixed morphology SNR, 
where the non-thermal radio shell surrounds a centrally peaked 
thermal X-ray emission \cite{mae02}. 
The X-ray plasma has 2 components, one at 1~keV 
and the other at 4 keV \cite{sak04}.
High abundances in the center of source indicate that 
part of emission is due to the heated SN ejecta. Most recent 
Chandra results on Sgr A East have shown evidences that one of the
sources of the region could be the kicked off NS from a SN II
explosion \cite{par05}. 
However the X-ray data show now that Sgr~A~East,
apart from the high plasma temperature and 
from beeing in expantion against a very dense medium,
is not an exceptional SNR. It appears to be the product of
a typical SN II or a SN Ia occured about 10$^4$~yr. 
Assuming a certain distance
of the SN from Sgr~A$^*$, the shell of swept up ISM 
could have reached the SMBH feeding it and triggerig 
a Sgr~A$^*$ outburst of hard emission, later reflected by Sgr~B2 \cite{mae02}.
\subsection{The recent results}
The years 2003 and 2004 have seen several new developments  
in the domain. In particular a series of large multiwavelength campaigns
have been performed in order to obtain broad band measures
on the variable emission from Sgr~A$^*$.
In 2003, the VLT \cite{gen03}, followed by
the Keck \cite{ghe05}, could reveal that Sgr~A$^*$ 
is flaring also in the NIR band. 
The NIR flares appear more frequent (several / day)
than the X-ray ones ($\sim$ 1 /day)
and their red spectra extending in the MIR domain \cite{eis05,ghe04,cle05} 
confirm that synchrotron is the dominant IR radiation mechanism.
The IR observations also provided the spectacular evidence 
that the emission appears modulated with a period of 17~mn \cite{gen03}.
If such a period is associated to the last stable orbit 
of an accretion disk such timescale implies that the SMBH 
is rotating at 50$\%$ of the maximum allowed spin.
Revisiting X-ray flare variability Aschebach et al. \cite{asc04} 
found a serie of possible quasi periods in the power spectra.
These periods appear in relation to the characteristic 
frequencies (keplerian, vertical and radial oscillations) 
of a disk orbiting around a BH with maximum spin. 
Althought these results are extremely exciting
they are still rather controversial and 
the reported X-ray periodicities from Sgr~A$^*$
need to be confirmed by more significant measurements.
Firm detections of QPO from Sgr~A$^*$ with periods in the 
range 15-30 mn would certainly favor the accretion disk models 
and would provide  
strong constraints on the mass an spin of the SMBH at the GC.
The first flare simultaneously observed in IR and X-rays 
was detected using Chandra and the VLT \cite{eck04}.
The flare was however very weak and a much stronger
event was observed simultaneously with XMM and the HST
during the large 2004 multiwavelength observation campaign 
of Sgr~A$^*$ \cite{yus05}. This campaign based on a
XMM-Newton large project 
involved radio (VLA, ATCA), sub-mm (CSO, SMT, NMA, BIMA), 
IR (VLT, HST) and gamma-ray (INTEGRAL, HESS) observatories.
Two bright (factor 35) X-ray flares were observed with 
XMM-Newton (Fig.~\ref{fig:Goldwurm-fig3}) \cite{bel05a}
and the September one could be observed with the NICMOS camera
of the HST. 
The NIR and X-ray flares are very similar in shape 
and the lack of time lags and the measured spectral slopes
may indicate that X-rays are indeed produced by inverse 
compton scattering of the same electrons that produce 
the NIR synchrotron emission off the sub-mm-radiation \cite{yus05}.

\section{The Galactic Center in gamma-rays}
\begin{figure}[htb]
\begin{center}
\epsfig{file=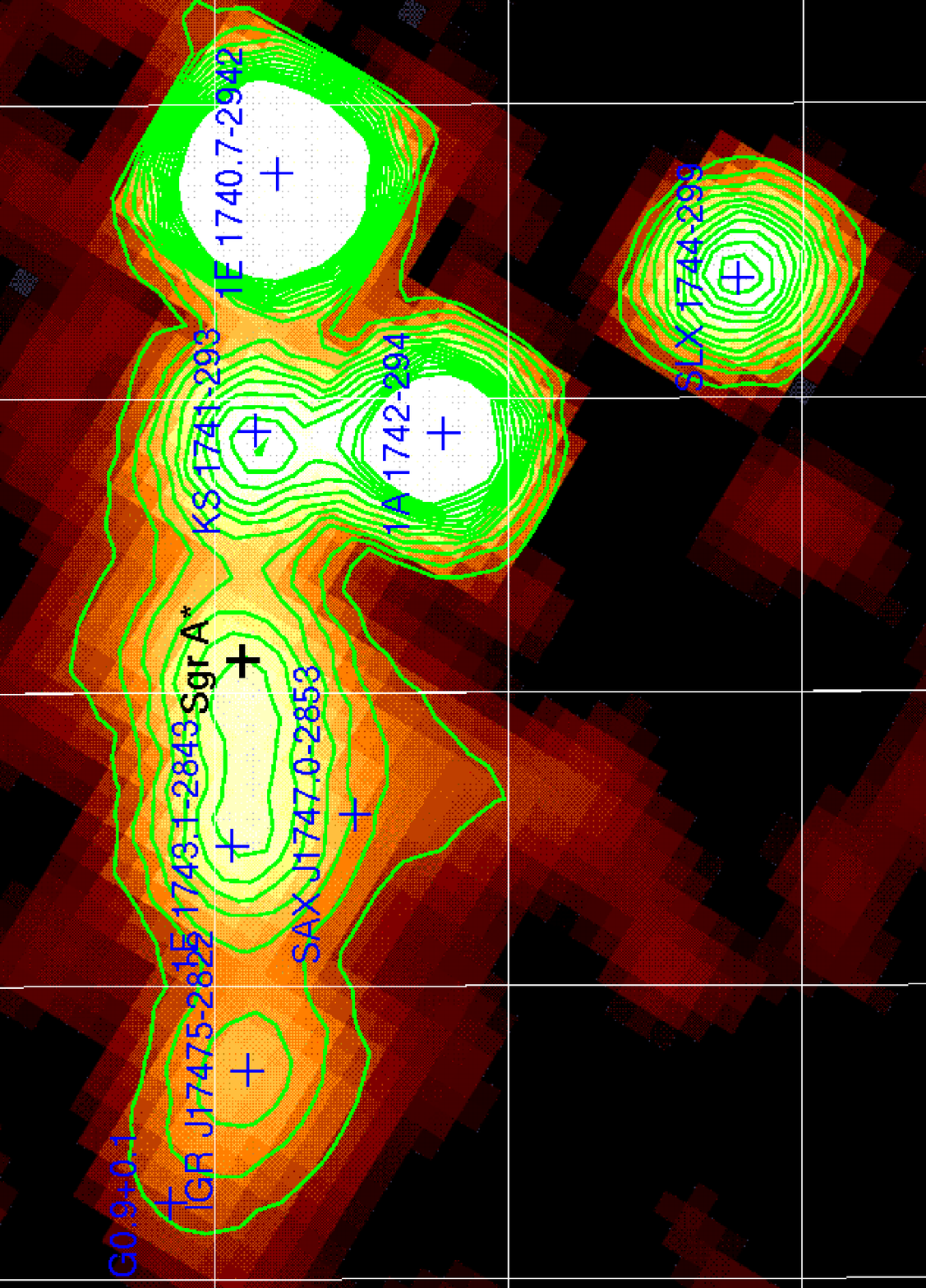,angle=-90, width=5cm}~~
\epsfig{file=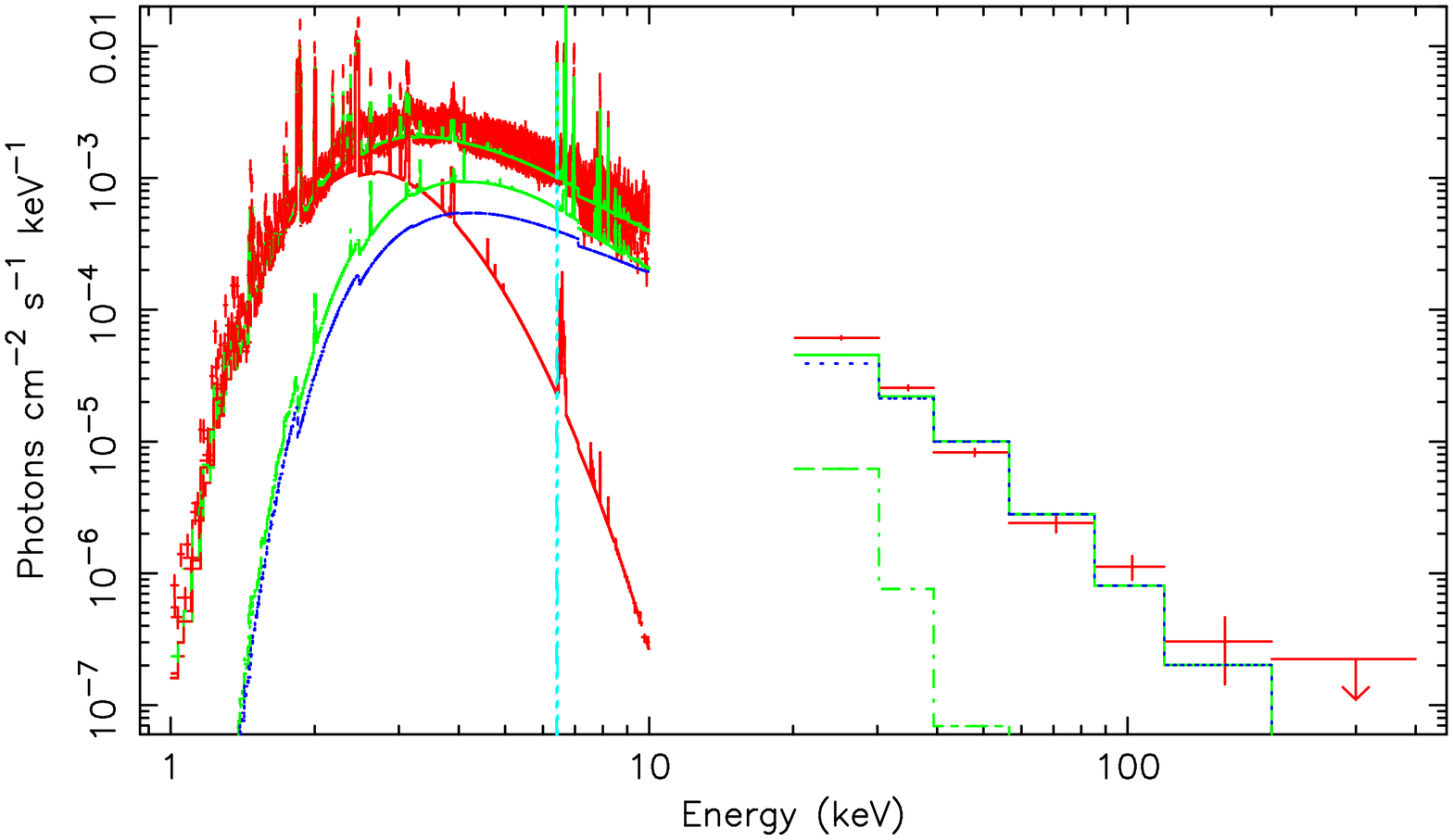,angle=-90, width=6.3cm}
\caption{The 20-40~keV INTEGRAL/IBIS images of the GC showing the 
source in Sgr~A (left) and
combined XMM-IBIS spectrum for this excess (right)}
\label{fig:Goldwurm-fig2}
\end{center}
\end{figure}

\begin{figure}[htb]
\begin{center}
\epsfig{file=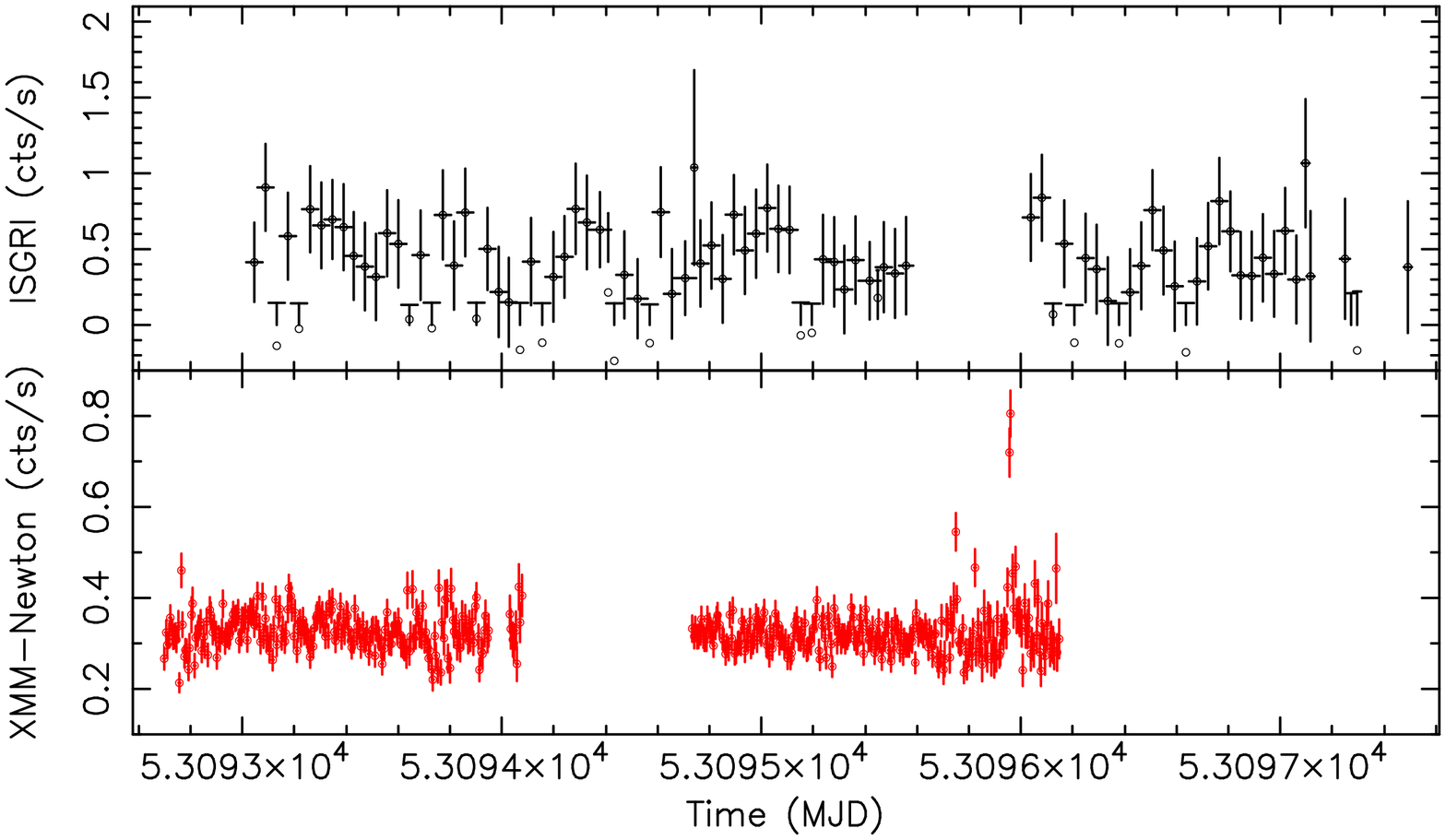,angle=-90, width=5.5cm}~~~~
\epsfig{file=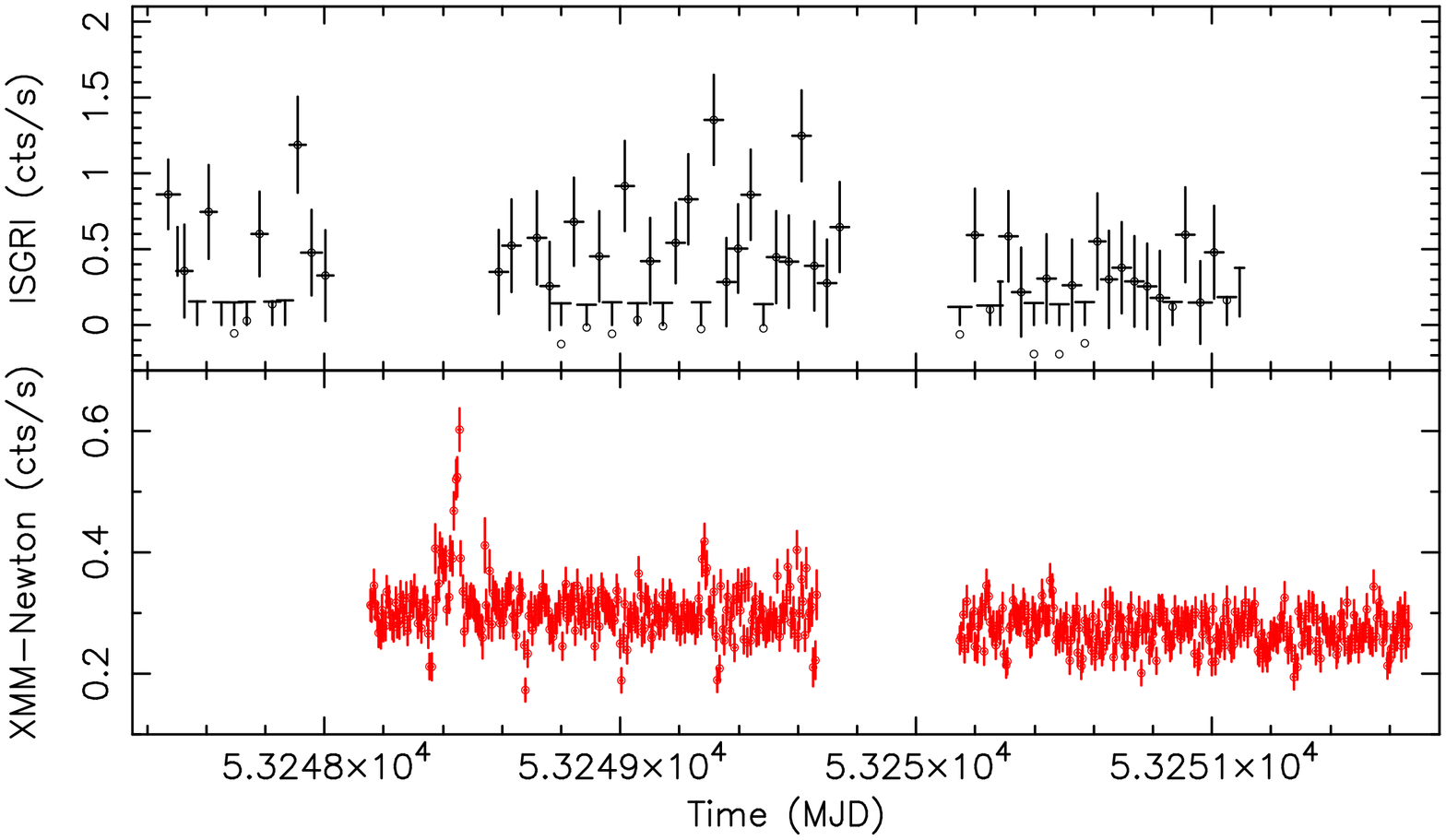,angle=-90, width=5.5cm}
\caption{
The 20-30~keV INTEGRAL/IBIS ligth curves (black) of the central source
in March (left) and Sep. (right) 2004. The
2 Sgr~A$^*$ flares in the XMM 2-10~keV light curves (red) 
occur during INTEGRAL radiation belt passages. 
}
\label{fig:Goldwurm-fig3}
\end{center}
\end{figure}

The INTEGRAL observatory monitored, with the IBIS/ISGRI telescope, 
the GC region for more than 7~Ms between 2003 and 2004, 
obtaining, with an effective exposure of 4.7~Ms,
the most precise images of the GC ever collected 
in the 20-600~keV band \cite{bel04, gol04, bel05b}.
In addition to the bright X-ray binaries INTEGRAL detected 
a faint and persistent high energy emission 
coming from the very center of the Galaxy (Fig.~\ref{fig:Goldwurm-fig2}),
compatible (within the 1$'$ error radius) with Sgr~A$^*$.
Due to the IBIS angular resolution ($\sim$ 13$'$ FWHM) this source
(IGR~J17456--2901) cannot be clearly associated to the SMBH or 
to other objects of the dense central region.
The lack of variability and of a bright discrete X-ray counterpart
suggest that it is rather a compact and yet diffuse emission. 
The INTEGRAL spectrum was compared to the 1-10~keV one obtained from
XMM-Newton (partly simultaneous) data integrating over the region
of the IBIS point spread function \cite{bel05b}.
The spectra combine well (Fig.~\ref{fig:Goldwurm-fig2}),
but the thermal plasma with kT of 8 keV used to model the bulk of the
X-ray diffuse emission cannot explain the data at $>20$~keV,
neither can the contributions of the transient point sources seen by Chandra
or XMM. 
A non-thermal component extending up to 120-200 keV with spectral slope of
photon index 3 is clearly present and its origin is still unexplained.   
Simultaneous XMM-INTEGRAL observations performed during the 2004 campaign
are not conclusive on the possible detection with INTEGRAL 
of the Sgr~A$^*$ X-ray flares since 
the 2 events observed with XMM occured during the INTEGRAL passage 
into the radiation belts (Fig.~\ref{fig:Goldwurm-fig3}).
However, even if the Sgr A* X-ray flares extend at $>$ 20 keV
with their hard slope, they are too sparse to fully account for 
the gamma-ray source.
In addition, INTEGRAL observed constant hard emission,
from Sgr B2  (IGR~J17475-2822 in Fig.~\ref{fig:Goldwurm-fig2}). 
This strongly confirm the thesis of a reflection nebula for Sgr~B2 \cite{rev04}.
%
%
%
%

\begin{figure}[htb]
\begin{center}
\epsfig{file=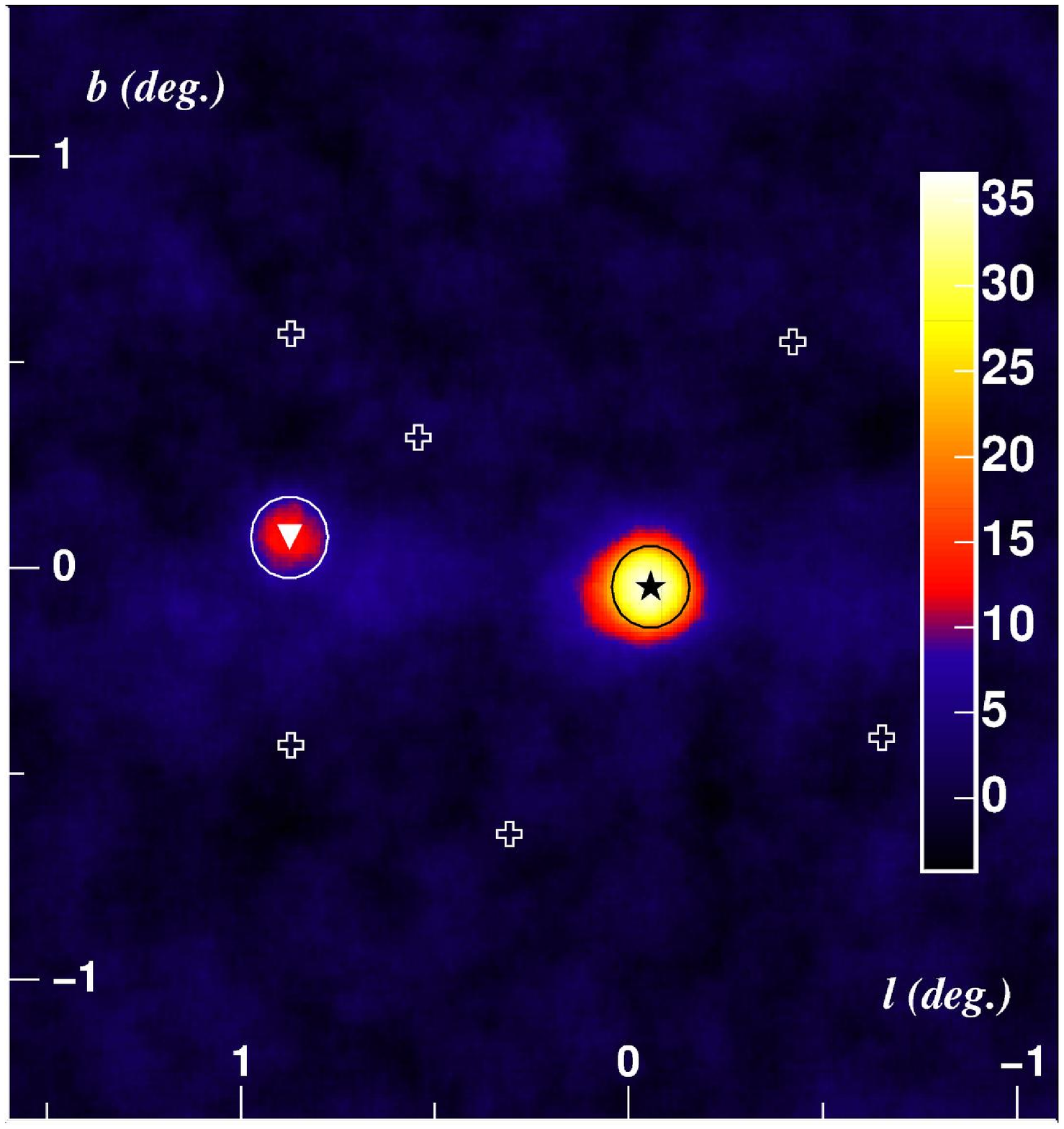, width=4cm}~~~~~~
\epsfig{file=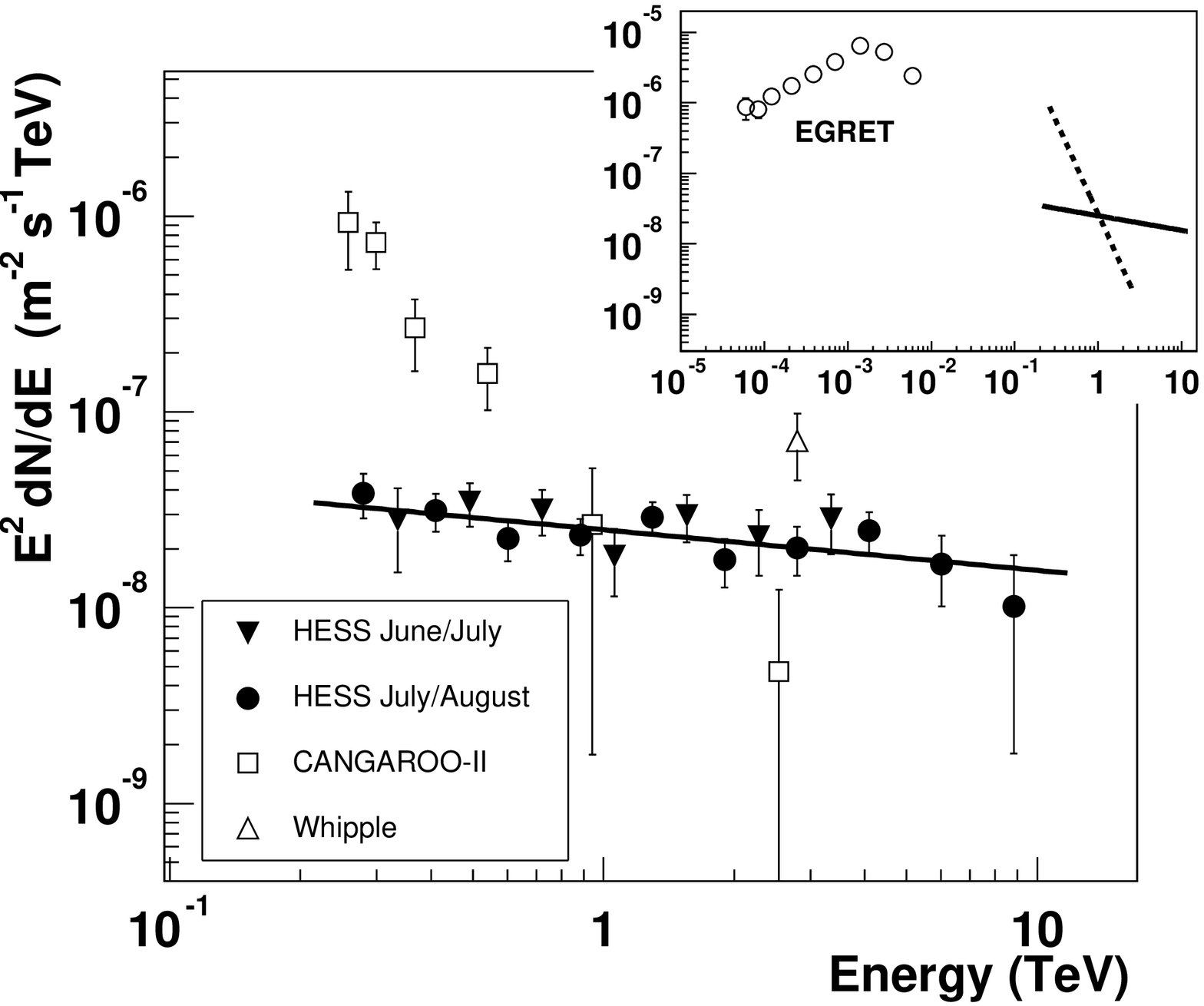, width=5cm}
\caption{HESS image of the GC showing the source in Sgr~A (star)
(Aharonian et al. 2005 {\it A}\&{\it A} 432 L25) (left) 
and the HESS spectrum of this source (right).
}
\label{fig:Goldwurm-fig4}
\end{center}
\end{figure}

IGR~J17456--2901 could be linked to the VHE gamma-ray emission 
observed by several Atmospheric Cherenkov Detectors. 
HESS, the most sensitive and precise of them, reported the presence of
a TeV source centered within 1$'$ from Sgr~A$^*$ \cite{aha04} 
(Fig.~\ref{fig:Goldwurm-fig4}).
The source is constant 
and display power-law spectrum extending from 300 GeV up to 10 TeV.
This emission cannot be explained by heavy dark matter particle 
annihilation and is probably due to interactions of particles accelerated
at very high energies. 
However the mechanism and site of acceleration,
the expanding shell of the Sgr~A~East SNR \cite{cro05} or the 
regions close to the SMBH horizon \cite{ahaner05}, are not yet identified.
The EGRET source observed between 50 MeV and 10 GeV 
(3EG~J1746-2852) located at 0.2$^{\circ}$ from Sgr~A$^*$ \cite{may98,har99}
seems too far to be the 1~GeV counterpart for the INTEGRAL
and the HESS sources, but in this complex region the EGRET data 
are not conclusive. 
Gamma-ray emission from the GC has now been clearly detected 
but its origin and nature are not yet understood.

The Chandra, XMM-Newton, INTEGRAL and HESS monitoring of the GC
will continue in the coming months/years, hopefully
coupled to NIR, sub-mm and radio correlated observing programs. 
These programs will possibly settle the issue of periodicities
in the X-ray flares and will provide measures of the broad band spectra
of the Sgr~A$^*$ flares.
Solving the puzzle of the hard X-ray emission will however necessitate 
focusing instruments in this energy domain, as Simbol--X expected to fly 
at the beginning of the next decade \cite{fer05}.
In the near future GLAST will probably unveil the mystery of the 
EGRET source at the GC and the next generation of ACD detectors 
will map the region at TeV energies with increased precision.
%
%
%
\begin{small}

\end{small}
%
\label{GoldwurmEnd}
\end{document}